\documentstyle[sprocl,epsf]{article}

\def\be{\begin{equation}}
\def\ee{\end{equation}}
\def\bea{\begin{eqnarray}}
\def\eea{\end{eqnarray}}

\begin{document}

\title{BOSE-EINSTEIN FINAL STATE SYMMETRIZATION
FOR EVENT GENERATORS OF HEAVY ION COLLISIONS}

\author{U.A. Wiedemann$^{a,b}$, J. Ellis$^c$, U. Heinz$^a$ and K. Geiger$^d$}

\address{$^a$Institut f\"ur Theoretische Physik, Universit\"at Regensburg,
D-93040 Regensburg\\
$^b$Physics Department, Columbia University, New York, N.Y. 10027, U.S.A.\\
$^c$Theoretical Physics Division, CERN, CH-1211 Geneva 23\\
$^d$Physics Department, BNL, Upton, N.Y. 11973, U.S.A.}


\maketitle
\abstracts{ 
We discuss algorithms which allow to calculate identical 
two-particle correlations from numerical simulations of 
relativistic heavy ion collisions. A toy model is used to
illustrate their properties.}

\section{Introduction}
The current relativistic heavy ion program at CERN and BNL aims
at investigating the equilibration properties of hadronic matter
at extreme temperatures and densities where quarks and gluons
are expected to be the physically relevant degrees of freedom
for particle production processes. The theoretical discussion
of these collision systems is complicated by their mesoscopic
character. They are not sufficiently small to allow for an 
analytical description in terms of elementary processes. They
are not sufficiently large to take a description in terms of
macroscopic observables for granted. Even if simple 
thermodynamically or hydrodynamically inspired models account
for the data, the task remains to understand the microscopic
origin of their success, and to establish to what extent
such an agreement is necessary or accidental.

Numerical simulations of relativistic heavy
ion collisions are an adequate tool to attack these questions.
They allow to propagate sufficiently many degrees of freedom
from some initial condition, and they thus test the
consequences of different assumptions about the microscopic
in medium dynamics. Many event generators have been developed
to this aim~\cite{YP97}. Irrespective of their particular model 
assumptions, they return for each simulated event a set of $N_m$ phase
space emission points $\lbrace (\check{\bf r}_i,\check{\bf p}_i,
\check{t}_i)\rbrace_{i\in [1,N_m]}$ together with
information, e.g., about the identity (PID) of the
particle $i$, the parent resonance it originates from, the
number of rescatterings it has undergone, etc.

The major aim of event generator studies is to reduce step by 
step the class(es) of dynamical scenarios consistent with 
experimental data, thus obtaining an increasingly better
picture of the physics. To this 
end, one mainly compares the simulated particle ratios and
one-particle momentum spectra to experiment. However, one
can imagine different dynamical scenarios which all account
for the same simulated momentum distributions 
$\lbrace{(\check{\bf p}_i)}\rbrace_{i\in [1,N_m]}$ but differ 
with respect to the simulated space-time structure 
$\lbrace{(\check{\bf r}_i,\check{t}_i)}\rbrace_{i\in [1,N_m]}$.
Identical two-particle correlations $C({\bf q},{\bf K})$,
here written as a function of the measured pair momentum
$K = \textstyle{1\over 2}(p_1 + p_2)$
and relative pair momentum $q = p_1 - p_2$
are sensitive to the latter and can thus play an important
complementary role in specifying the parameter space of event 
generators. Here, we discuss algorithms of the type
  \begin{equation}
    \Bigg\lbrace
    \lbrace (\check{\bf r}_i, \check{\bf p}_i, 
    \check{t}_i)\rbrace_{i\in [1,N_m]}
    \Bigg\rbrace_{m\in [1,N_{\rm ev}]}
    \Longrightarrow C({\bf q},{\bf K})\, .
    \label{eq1}
  \end{equation}
For notational convenience, we restrict the following discussion 
to one particle species only, like-sign pions say. Also, we neglect
final state Coulomb and strong interactions~\cite{AHR97}. The set
$\lbrace{(\check{\bf r}_i,\check{\bf p}_i,
\check{t}_i)}\rbrace_{i\in [1,N_m]}$
denotes then the phase space emission points of the $N_m$
like-sign pions generated in the $m$-th simulated event. 
The event generator simulates thus a classical phase space 
distribution
 \begin{equation}
 \label{eq2}
   \rho_{\rm class}({\bf r},{\bf p},t) =
   {1\over N_{\rm ev}} \sum_{m=1}^{N_{\rm ev}} \sum_{i=1}^{N_m}
   \delta^{(3)}({\bf r}-\check{\bf r}_i)\,
   \delta^{(3)}({\bf p}-\check{\bf p}_i)\, 
   \delta(t-\check{t}_i)\, .
 \end{equation}
%

\section{From phase space densities to momentum correlations}
\label{sec2}

The standard particle interferometric analysis of heavy ion collisions
starts from the relation between the measured momentum correlations
of identical particles and the quantum mechanical Wigner phase
space density $S(x,K)$ of particle emitting sources in the collision
region~\cite{He96},
  \begin{equation}
         C({\bf q},{\bf K}) = {\cal N} \left(1 + 
     {\left\vert \int d^4x\, S(x,K)\, e^{iq{\cdot}x}\right\vert^2 
      \over
       \int d^4x\, S(x,p_1)\, \,\int d^4y\, S(y,p_2)  }\right)\, .
   \label{eq3}
  \end{equation}
This expression can be generalized to include final state 
interactions~\cite{AHR97}. Depending on the definition of 
the two-particle correlator in terms of single- and 
two-particle cross sections~\cite{MV97,ZSH98} (respectively
depending on its construction from experimental data), the 
normalization is either ${\cal N} = 1$ or ${\cal N} < 1$.
The calculation of deviations of ${\cal N}$ from $1$ typically
involves multiparticle symmetrization effects~\cite{W98,ZSH98},
which we do not review here. Our discussion will thus be for
${\cal N} = 1$.

In the light of Eq.~(\ref{eq3}), calculating the two-particle
correlator $C({\bf q},{\bf K})$ from an event generator output
amounts to specifying the emission function $S(x,K)$ in terms
of the generated phase space points
$\lbrace{(\check{\bf r}_i,\check{\bf p}_i,
\check{t}_i)}\rbrace_{i\in [1,N_m]}$, i.e., it amounts to 
an interpretation of $(\check{\bf r}_i, \check{\bf p}_i, \check{t}_i)$. 
In sections~\ref{sec2a} and~\ref{sec2b}, we focus on two different
interpretations referred to as {\it quantum mechanical} and 
{\it classical}, though they are both conceptually on an equal 
footing. Before doing that, we list here three requirements
which an algorithm implementing (\ref{eq1})  should fulfill:
  \begin{enumerate}
    \item
    Event generators use {\it probabilities} to describe the
    collision process. Bose-Einstein correlations are obtained
    from squaring production {\it amplitudes}. Hence, the
    generated momenta $\check{\bf p}_i$, $\check{\bf p}_j$
    do not show the enhancement at low relative momenta characteristic
    for Bose-Einstein final state symmetrization. The
    prescriptions (\ref{eq1}) should calculate this effect
    a posteriori.
    \item
    The strength of Bose-Einstein correlations depends on the
    distance of the identical particles in phase space, not
    in momentum space. We thus require the prescriptions
    (\ref{eq1}) to use the entire phase space information,
    and not only the generated momentum information.
    `Weighting' or `shifting' prescriptions which are based only on the
    latter may successfully match the measured momentum
    correlations but obviously do not allow to test the
    simulated space-time structure.
    \item
    In general, Bose-Einstein statistics can affect particle
    multiplicity distributions during the particle production
    process. However, event generators are tuned to reproduce
    the average particle multiplicities in agreement with the
    data. We require hence that the prescriptions (\ref{eq1})
    conserve particle multiplicities. If the correlator is
    interpreted as a factor, relating measured two-particle
    differential cross sections to those simulated by the
    event generator, $d^6\sigma_{\rm meas}/d^3{\bf p}_1\, d^3{\bf p}_2$
    $= C({\bf q},{\bf K})$  $
    d^6\sigma_{\rm sim}/d^3{\bf p}_1\, d^3{\bf p}_2$, then
    this implies~\cite{LS95,W98} that ${\cal N} < 1$ in (\ref{eq3}).
    This however plays no role in what follows
    since the space-time analysis of correlation data can be based
    entirely on the momentum dependence of $C({\bf q},{\bf K})$,
    disregarding its absolute normalization.
  \end{enumerate}
%
\subsection{``Quantum'' interpretation of event generator output}
\label{sec2a}

In the {\it quantum mechanical} 
interpretation~\cite{Weal97,ZWSH97,W98,EGHW98}, the set of phase 
space points $\lbrace \lbrace (\check{\bf r}_i, \check{\bf p}_i, 
\check{t}_i)\rbrace_{i\in [1,N_m]} \rbrace_{m\in [1,N_{\rm ev}]}$
is associated
to the centers of Gaussian one-particle wavepackets according 
to~\cite{MP97,Weal97,ZWSH97} 
  \begin{eqnarray}
    (\check{\bf p}_i, \check{\bf r}_i, \check{t}_i) \longrightarrow
    f_i({\bf x},\check{t}_i) 
    = {1\over (\pi\sigma^2)^{\footnotesize{3/4}} }
    {\rm e}^{-\textstyle{1\over 2\sigma^2}\, 
      ({\bf x}-\check{\bf r}_i)^2 + i\check{\bf p}_i\cdot {\bf x} }\, .
    \label{eq4}
  \end{eqnarray}
These wavepackets $f_i$ are quantum mechanically best localized
states, i.e., they saturate the Heisenberg uncertainty relation
with $\Delta x_i = \sigma/\sqrt{2}$ and 
$\Delta p_i = 1/\sqrt{2}\sigma$.

Taking only two-particle symmetrized contributions into account,
i.e., adopting the so-called pair approximation, all spectra can
be written in terms of the functions $s_i({\bf p})$,
which coincide with the corresponding one-particle spectrum for
a source emitting only one particle at the phase space point $i$.
For the one- and two-particle correlator one 
finds~\cite{ZWSH97,Weal97}
 \begin{eqnarray}
   E_p {d^3N\over d^3p} &=& {1 \over N_{\rm ev}} \sum_{m=1}^{N_{\rm ev}}
   \nu_m({\bf p}) = {1 \over N_{\rm ev}} 
   \sum_{m=1}^{N_{\rm ev}} \sum_{i=1}^{N_m} s_i({\bf p})\, ,
   \label{eq5} \\
    C({\bf q},{\bf K}) &=&  1 + e^{-\sigma^2 {\bf q}^2/2} 
     { \sum_{m=1}^{N_{\rm ev}} \left[ 
       \left\vert \sum_{i=1}^{N_m} s_i({\bf K})\,
         e^{iq^0\check{t}_i - i{\bf q}\cdot \check{\bf r}_i} \right\vert^2
      - \sum_{i=1}^{N_m} s_i^2({\bf K}) \right]
      \over 
        \sum_{m=1}^{N_{\rm ev}} \left[ \nu_m({\bf p}_a)\, \nu_m({\bf p}_b) 
      - \sum_{i=1}^{N_m} s_i({\bf p}_a)\,s_i({\bf p}_b) \right]}\, ,
 \label{eq6} \\
   s_i({\bf p}) &=& \pi^{-3/2}\, \sigma^3\, 
   e^{ - \sigma^2 ({\bf p}-\check{\bf p}_i)^2}\, .
 \label{eq7}
 \end{eqnarray}
Here, the terms subtracted in the numerator and denominator of 
$C({\bf q},{\bf K})$ are finite multiplicity corrections which
become negligible for large particle multiplicities~\cite{Weal97}.
This algorithm is consistent with an emission function $S(x,{\bf K})$ 
  \begin{eqnarray}
    S(x,{\bf K}) &=& \int d\check{\bf r}_i\, d\check{\bf p}_i\, d\check{t_i}\,
    \rho_{\rm class}(\check{\bf r}_i,\check{\bf p}_i,\check{t_i})\, 
    s_0({\bf x}-\check{\bf r}_i,t-\check{t_i},{\bf K}-\check{\bf p}_i)\, ,
    \label{eq8}\\
    s_0(x,{\bf K}) &=&
    \textstyle{1\over \pi^3}\, \delta(t)\,
    {\rm e}^{-\textstyle{1\over \sigma^2}{\bf x}^2
    -\textstyle{\sigma^2}{\bf K}^2}\, ,
    \label{eq9}
  \end{eqnarray}
which is given by a folding relation between a classical distribution 
$\rho_{\rm class}$ of wavepacket centers and the elementary source 
Wigner function $s_0(x,{\bf K})$. The latter has at emission an 
optimal quantum mechanical phase space localization around phase 
space points $(\check{\bf r}_i, \check{\bf p}_i, \check{t}_i)$ 
with spatial uncertainty $\sigma$ and momentum uncertainty $1/\sigma$. 

In the algorithm (\ref{eq5})-(\ref{eq7}), the spectra
are discrete functions of the input $(\check{\bf r}_i, \check{\bf p}_i, 
\check{t}_i)$ but they are continuous in the observable momenta 
${\bf p}_1$, ${\bf p}_2$ and hence, no binning is necessary. Also, 
the number of terms to be calculated in (\ref{eq5}) and (\ref{eq6})
grows linear in $N$, not quadratic as in other prescriptions.
However, both the one- and two-particle spectra depend in this
algorithm on the wavepacket width $\sigma$. The role of this
parameter will be discussed 
in the context of our toy model study in section~\ref{sec3}.

\subsection{``Classical'' interpretation of event generator output}
\label{sec2b}
%
In the {\it classical} interpretation~\cite{ZWSH97,EGHW98}, 
the phase space points $(\check{\bf r}_i, \check{\bf p}_i, \check{t}_i)$ 
are seen as a discrete approximation of the on-shell Wigner phase 
space density $S(x,p)$, 
  \begin{equation}
    S(x,{\bf p}) = \rho_{\rm class}({\bf x},{\bf p},t)\, .
    \label{eq10}
  \end{equation}
This defines both the two-particle correlator via (\ref{eq3}) and the
one-particle spectrum via the $x$-integration over (\ref{eq10}).
To simplify a numerical implementation of this calculation, it is 
useful to replace the $\delta$ functions of $\rho_{\rm class}$
in momentum space by rectangular `bin functions'~\cite{ZWSH97}, 
or by properly normalized Gaussians~\cite{EGHW98} of width $\epsilon$,
  \begin{equation}
    \delta^{(\epsilon)}_{\check{\bf p}_i,{\bf p}}
    = {1\over (\pi\epsilon^2)^{3/2}}
      \exp \left( - (\check{\bf p}_i - {\bf p})^2/\epsilon^2\right)\, .
    \label{eq11}
  \end{equation}
In the limit $\epsilon \to 0$, these Gaussian bin functions reduce to
the properly normalized $\delta$-functions 
$\delta^{(3)}(\check{\bf p}_i - {\bf p})$ and we regain Eq. (\ref{eq2}).
The one-particle spectrum and two-particle correlator read 
then~\cite{ZWSH97,EGHW98}
  \begin{equation}
     E_p {d^3N\over d^3p} = \int d^4x\, S(x,{\bf p}) =
       {1 \over N_{\rm ev}} \sum_{m=1}^{N_{\rm ev}}
       \sum_{i=1}^{N_m} \delta^{(\epsilon)}_{\check{\bf p}_i,{\bf p}}\, ,
   \label{eq12} 
  \end{equation}
  \begin{equation}
   C({\bf q},{\bf K}) = 1 + { \sum_{m=1}^{N_{\rm ev}} \left[
   \left\vert \sum_{i=1}^{N_m} 
       \delta^{(\epsilon)}_{\check{\bf p}_i,{\bf K}}\, 
       e^{iq^0\check{t}_i - i{\bf q}\cdot \check{\bf r}_i} \right\vert^2 
        - \sum_{i=1}^{N_m} \left(\delta^{(\epsilon)}_{\check{\bf p}_i,{\bf K}}
              \right)^2
   \right]
   \over
   \sum_{m=1}^{N_{\rm ev}} \left[ \left(\sum_{i=1}^{N_m} 
       \delta^{(\epsilon)}_{\check{\bf p}_i,{\bf p}_1}\right)
       \left(\sum_{j=1}^{N_m} 
         \delta^{(\epsilon)}_{\check{\bf p}_j,{\bf p}_2}\right)
   - \sum_{i=1}^{N_m} \delta^{(\epsilon)}_{\check{\bf p}_i,{\bf p}_1}
                      \delta^{(\epsilon)}_{\check{\bf p}_i,{\bf p}_2}
   \right]}\, .
 \label{eq13}
 \end{equation}
The correlator (\ref{eq13}) is the discretized version of the
Fourier integrals in (\ref{eq3}).
The subtracted terms in the numerator and denominator
remove discretization errors which would amount to pairs constructed
from the same particles. The `classical' algorithm has the same
important advantages as the `quantum' one, it involves only $O(N_m)$
rather than $O(N_m^2)$ numerical manipulations per event, and its observables
are discrete functions of the input, but continuous functions of the
measured momenta; hence, no binning is necessary.  `Classical' and
`quantum' algorithm then differ only with respect to two points:
  \begin{enumerate}
    \item
      The classical algorithm has no analogue for the
      Gaussian prefactor $\exp\left( \right.$ 
      $\left. -\sigma^2\, {\bf q}^2/2\right)$
      in (\ref{eq6}) which is a genuine quantum effect stemming from
      the quantum mechanical localization properties of (\ref{eq4}).
    \item
      For the choice $\sigma = 1/\epsilon$, the bin functions
      $\delta^{(\epsilon)}_{\check{\bf p}_i,{\bf p}}$ is the
      classical counterpart of  
      the Gaussian single-particle distributions $s_i({\bf p})$.
      Finite event statistics puts a lower practical limit on $\epsilon$,
      but the physical momentum spectra are recovered in the limit 
      $\epsilon \to 0$, i.e., $\epsilon$ should be chosen as small
      as technically possible. In contrast, the quantum algorithm 
      treats $\sigma$ as finite
      physical localization of the single particle 
      wavepackets. The limit $\sigma \to \infty$ (which corresponds
      to $\epsilon \to 0$) is not the physically relevant case.
      It amounts according to (\ref{eq8}) to an emission function 
      which is infinitely extended in space, and hence~\cite{Weal97}
      $\lim_{\sigma\to\infty} C({\bf q},{\bf K}) = 1 + \delta_{{\bf q},0}$. 
  \end{enumerate}

\section{The Zajc model}\label{sec3}

Our final aim is to apply the above algorithms to event generators
of relativistic heavy ion collisions and high energy elementary
processes. A first step in this direction will be reported 
in~\cite{EGHW98}. Here, we illustrate the properties of these
algorithms for a toy model.

The strategy is to start from a simple expression
for the classical distribution $\rho_{\rm class}$, from which
the results of the algorithms of sections~\ref{sec2a} and ~\ref{sec2b}
can be calculated analytically. In a second step, we generate then
in a Monte Carlo procedure sets of phase space points
$\lbrace (\check{\bf r}_i,\check{\bf p}_i,\check{t}_i)\rbrace_{i\in[1,N_m]}$ 
according to this
distribution, and we test the numerical results against our exact 
analytical expressions. This allows us to make in an analytically
controlled setting statements about i) the $\sigma$-dependence of the
quantum algorithm, ii) the $\epsilon$-dependence of the
classical algorithm (especially: how small $\epsilon$ has to be
chosen to extract the limit $\epsilon \to 0$ numerically) 
and iii) the statistical requirements for the algorithms to work.

Here, we study these questions for the Zajc model which describes a 
position-momentum correlated phase space distribution of
emission points~\cite{Z93}
  \begin{eqnarray}
    \rho_{\rm class}^{\rm Zajc}({\bf r},{\bf p},t) &=& 
    {\cal N}_s\, \exp \left\{ -\frac{1}{2(1-s^2)} 
             \left(    \frac{{\bf r}^2}{R_0^2} 
                 -2s\frac{{\bf r}\cdot{\bf p}}{R_0 P_0}
                    +  \frac{{\bf p}^2}{P_0^2}
                 \right) \right\} \delta(t)\, ,
    \label{eq14} \\
    {\cal N}_s  &=& E_p\, 
    {N\over (2\pi)^3\, R_0^3\, P_0^3\, (1-s^2)^{3/2}} \, . 
    \label{eq15}
  \end{eqnarray}
This distribution is normalized such that the total
event multiplicity calculated from the one-particle spectrum
(\ref{eq12}) is $N$. The parameter $s$ smoothly interpolates 
between completely uncorrelated and completely position-momentum 
correlated sources. 
For $s \to 0$, the position-momentum correlation vanishes
and we are left with the product of two Gaussians in position
and momentum space. In the opposite limit
  \begin{equation}
  \label{eq16}
    \lim_{s\to 1}  \rho_{\rm class}^{\rm Zajc}({\bf r},{\bf p},t) 
       \sim \delta^{(3)}
    \left(\frac{{\bf r}}{R_0} - \frac{{\bf p}}{P_0}\right)\,\delta(t)\, ,
  \end{equation}
the position-momentum correlation is perfect. The corresponding
phase space volume of the distribution is proportional to 
  \begin{equation}
    V =  2^3\, R_0^3\, P_0^3\, (1-s^2)^{3/2}\, ,
    \label{eq17}
  \end{equation}
and goes to zero for $s \to 1$. This strong $s$-dependence 
allows to study the performance of our numerical algorithms 
for different phase space volumes. In the following subsections,
we discuss the $s$-dependence of the one-particle spectrum and 
two-particle correlator, focussing in sections~\ref{sec3a} and
~\ref{sec3b} on analytical results, and comparing these in 
section~\ref{sec3c} to numerical calculations.

\subsection{The Zajc model in the classical algorithm}
\label{sec3a}

Starting from an analytical emission function $S(x,{\bf p}) =$
$\rho_{\rm class}({\bf x},{\bf p},t)$ according to (\ref{eq10}),
the one-particle spectrum and two-particle correlator read 
  \begin{eqnarray}
     E_p{dN\over d^3p} &=& E_p {N\over (2\pi)^{3/2} P_0^3}\, 
                        \exp\left( -{ {\bf p}^2\over 2P_0^2}\right)\, ,
                        \label{eq18}\\
     C({\bf q},{\bf K}) &=& 1 + \exp 
     \Big( -{\bf q}^2 R_{\rm class}^2(\epsilon) \Big)\, ,
     \label{eq19}\\
    R_{\rm class}^2(\epsilon) &=& 
    R_0^2 \left( 1-s^2{1\over 1 + \epsilon^2/2P_0^2} 
                  - {1\over {(2 P_0 R_0)^2}}{1\over 1 + \epsilon^2/2P_0^2}
             \right) \, .
              \label{eq20}
  \end{eqnarray}
We recover the physical HBT radius from (\ref{eq20}) in the limit 
$\epsilon \to 0$ or by inserting (\ref{eq10}) directly into (\ref{eq2}).
The remarkable fact is that for sufficiently large $s$~\cite{Z93},
  \begin{equation}
    s > s_{\rm crit} = \sqrt{ 1 - {1\over (2R_0P_0)^2}}\, ,
    \label{eq21}
  \end{equation}
the HBT radius $R_{\rm class}^2$ becomes negative and the two-particle
correlator shows an unphysical rise of the 
correlation function with increasing ${\bf q}^2$. 
The change of sign in (\ref{eq19}) seems to be related to the violation
of the uncertainty relation by the emission function (\ref{eq14}):
In the limiting case (\ref{eq16}), $\rho_{\rm class}^{\rm Zajc}$
reduces to a perfect position-momentum correlated source which is
not consistent with the Heisenberg uncertainty relation. Also, the 
phase space volume $V$ in (\ref{eq17}) drops below 1 for 
$s > s_{\rm crit}$. In this sense, the
distribution $\rho_{\rm class}^{\rm Zajc}$ is a quantum mechanically
allowed emission function $S(x,{\bf p}) =$
$\rho_{\rm class}({\bf x},{\bf p},t)$ for $s < s_{\rm crit}$ only.

Clearly, an ansatz for a quantum mechanical Wigner phase space density 
$S(x,K)$ has to satisfy quantum mechanical localization properties. However,
realistic event generators satisfy this requirement anyway. 
The practical importance of (this limiting case of) the Zajc model
is thus that it provides an (extreme) testing ground for
our numerical algorithms. Analytically, we conclude already from
(\ref{eq20}) that a bin width sufficient to investigate the limit 
$\epsilon \to 0$, has to be small on the scale of the width of the 
generated momentum distribution,
  \begin{equation}
    \epsilon \ll \sqrt{2}\, P_0\, .
    \label{eq22}
  \end{equation}
At least for the present model study, this requirement is independent
of the strength of position-momentum correlations in the source.
%
\subsection{The Zajc model in the quantum algorithm}
\label{sec3b}
We now interprete the classical probability distribution
$\rho_{\rm class}^{\rm Zajc}$ of (\ref{eq14}) as a distribution
of phase space points associated with the centers of Gaussian
wavepackets according to (\ref{eq4}). The one-particle spectrum
and two-particle correlator are then readily calculated via the 
emission function (\ref{eq8}),
  \begin{eqnarray}
     E_p{dN\over d^3p} &=& E_p {N\over (2\pi)^{3/2} 
       (P_0^2 + 1/2\sigma^2)^{3/2}}\, 
         \exp\left( -{ {\bf p}^2\over 2P_0^2 + 1/\sigma^2}\right)\, ,
                        \label{eq23}\\
    C({\bf q},{\bf K}) &=& 1 + \exp \Big( -{\bf q}^2 R_{\rm qm}^2
                         \Big)\, ,
    \label{eq24} \\
    R_{\rm qm}^2 &=& {\sigma^2\over {1 + 2\sigma^2P_0^2}}
    \left( \sigma^2P_0^2 + R_0^2/\sigma^2 
           + 2R_0^2P_0^2(1-s^2) \right)\, .
    \label{eq25}
  \end{eqnarray}
In contrast to the classical algorithm, the radius parameter is now
always positive, irrespective of the value of $s$. Even if the classical 
distribution 
$\rho_{\rm class}^{\rm Zajc}(\check{\bf r},\check{\bf p},\check{t})$ 
is sharply localized in phase space, its folding with minimum uncertainty 
wave packets leads to a quantum mechanically allowed emission function 
$S(x,{\bf p})$ and to a correlator with falls off with increasing
${\bf q}^2$ as expected. However, the spread of the one-particle
momentum spectrum (\ref{eq23}) receives an additional contribution
$1/\sigma^2$. Choosing $\sigma$ too small increases this term beyond the
phenomenologically reasonable values, choosing it too large widens
the corresponding HBT radius parameters significantly. It was
argued~\cite{Weal97} that $\sigma$ can be interpreted as quantum
mechanical `size' of the particle, $\sigma \approx 1$ fm. Given 
the heuristic nature of these arguments and the significant
modifications this implies for the spectra (\ref{eq23}) and
(\ref{eq24}), it is however fair to say that presently $\sigma$
mainly plays the role of a regulator of unwanted violations of the 
quantum mechanical uncertainty principle while a deeper understanding
of its origin in the particle production dynamics is still missing.
%
\begin{figure}[ht]\epsfxsize=9.5cm 
\centerline{\epsfbox{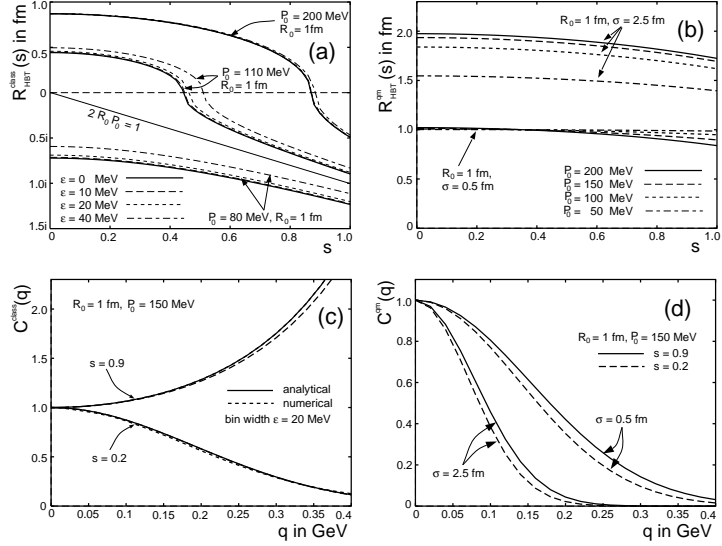}}
\caption{
Generic properties of the one-dimensional Zajc model. (a) HBT-radius parameter
(\ref{eq20}) of the classical interpretation as a function of
the position momentum correlation $s$. The plot shows the
HBT radius for different combinations of the model-parameters
$R_0$ and $P_0$, and for different sizes of the bin width $\epsilon$
needed in numerical implementations. (b) same as (a) for
the quantum version of the model, eq. (\ref{eq25}). The dependence
of the HBT radius on the wavepacket width $\sigma$ is seen clearly.
(c) and (d) The two-particle correlator in the classical (c) and
quantum mechanical (d) version of the model for different sets
of model parameters. The numerical results are obtained for $50$
events of multiplicity $1000$, and agree very well with the analytical
calculations.
}\label{fig1}
\end{figure}
%
\subsection{Numerical simulation in the Zajc model}
\label{sec3c}
We have studied the performance of our Bose-Einstein algorithms
by generating with a random number generator sets
$\lbrace{(\check{\bf r}_i,\check{\bf p}_i,
\check{t}_i)}\rbrace_{i\in [1,N_m]}$ of phase space points
according to the
distribution $\rho_{\rm class}^{\rm Zajc}$ and comparing the
numerical results of our algorithms to the analytical expressions
of section~\ref{sec3a} and~\ref{sec3b}. 

In Fig.~\ref{fig1}(a), the $\epsilon$-dependence of the HBT-radius 
parameter (\ref{eq20}) is depicted. For fixed bin width $\epsilon$,
the approximation of the true HBT radius parameter by 
$R_{\rm class}^2(\epsilon)$ is seen to become better with 
increasing $P_0$, in agreement with (\ref{eq22}).
For the HBT radius obtained from the quantum version of the
Zajc model and depicted in Fig.~\ref{fig1}(b), the situation is 
both qualitatively and quantitatively different. Now, the
HBT radius is always positive, since
the Gaussian wavepackets take quantum mechanical localization
properties automatically into account. Also, the $s$-dependence
of the radius is seen to be much smaller, since the wavepackets
smear out the unphysically strong position-momentum correlations
present in the classical distribution
$\rho_{\rm class}^{\rm Zajc}$. The HBT radius depends
not only on the geometrical size $R_0$, and on the momentum
width $P_0$ of the source, but also on the wavepacket width 
$\sigma$. As seen in Fig.~\ref{fig1}(b), this wavepacket
width affects the HBT radius 
and its $P_0$-dependence significantly for $\sigma > R_0$.

In Fig.~\ref{fig1}(c) and (d) we have presented the two-particle
correlation functions corresponding to these HBT-radius parameters
for characteristic values of the model parameters. The analytical
results, obtained by plotting (\ref{eq19}) and (\ref{eq24}) are
compared to the results of an application of the event generator
algorithms (\ref{eq6}) and (\ref{eq13}) for a Monte Carlo phase 
space distribution $\rho^{\rm Zajc}_{\rm class}$ of phase space
points. The present plot is obtained for $50$ events
of multiplicity $1000$. 
The differences between
analytical and numerical results have their origin in statistical
fluctuations and become smaller with increasing number of events
$N_{\rm ev}$ or event multiplicity $N_m$. The good
agreement between analytical and numerical results in 
Fig.~\ref{fig1}(c,d) indicates the relatively low 
statistical requirements of our algorithms. The reason is
that both algorithms associate to the {\it discrete}
set of generated momenta $\check{\bf p}_i$ {\it continuous}
functions of the measured momenta ${\bf p}_1$, ${\bf p}_2$.
This smoothens any statistical fluctuation significantly.
For the classical algorithm, small deviations between numerical
and analytical results are still visible in Fig.~\ref{fig1}(c),
while the results of the quantum algorithm coincide
within line width. This can be traced back to the Gaussian
prefactor $\exp\left( -\sigma^2\, {\bf q}^2/2\right)$
in eq. (\ref{eq6}) which provides an additional smoothening
of statistical fluctuations not present in the classical algorithm.

\section*{Acknowledgments}
This is a report on work in progress. The presenting author,
U.A.W., would like to acknowledge a DFG-Habilitationsstipendium.
He also thanks the CRIS'98 Local Organizing Comittee for generous
support which made his participation at this Conference possible.

\end{document}